\documentclass[preprint, showpacs, superscriptaddress, pre]{revtex4-2}
\usepackage{amsmath}
\usepackage{amssymb}
\usepackage{graphicx}
\newcommand{\be}{\begin{equation}}
\newcommand{\ee}{\end{equation}}
\newcommand{\FI}{f_w}
\newcommand{\U}{U}

\newcommand{\gedi}{GEDI}
\newcommand{\gi}{G_{\rm in}}

\newcommand{\lam}{\lambda}
\newcommand{\mubar}{\overline{\mu}}

\begin{document}

\title{Statistical  physics of an asset exchange model with  investment  and guaranteed income}

\author{J. Tobochnik}
\email{jan\_tobochnik@brown.edu}
\affiliation{Department of Physics, Brown University, Providence, Rhode Island 02912}
\affiliation{Department of Physics, Kalamazoo College, Kalamazoo, Michigan 49006}

\author{Harvey Gould}
\affiliation{Department of Physics, Boston University, Boston, Massachusetts 02215}
\affiliation{Department of Physics, Clark University, Worcester, Massachusetts 01610}

\author{W. Klein}
\affiliation{Department of Physics, Boston University, Boston, Massachusetts 02215}
\affiliation{Center for Computational Science, Boston University, Boston, Massachusetts 02215}

\date{\today}
\keywords{asset exchange, investment, income, equality, driven system}

\begin{abstract}
An agent-based model of the economy is generalized to incorporate investment and guaranteed income mechanisms in addition to the   exchange and distribution mechanisms  considered in an earlier model. We use the tools of statistical physics to show that the system is effectively ergodic, is not in  equilibrium, but   reaches a steady state with occasional large fluctuations because of the effects of multiplicative noise from the investment mechanism.  We find realistic wealth distributions and realistic values of the Gini coefficients and the Pareto index. 
\end{abstract}

\pacs{}

\maketitle

\section{Introduction} 

The concepts and techniques of statistical physics have been successful in treating many-body systems even when the components are not  simple objects. For example, statistical physics has been widely applied to biological and social systems. Most of these  systems are not in thermal equilibrium. An important question is what feature of these systems are primarily responsible for preventing these systems from reaching  equilibrium. We will see that the presence of multiplication noise is one such feature.

Many complex systems have been explored using agent based models. For example, these models have been successful at providing insight into why wealth inequality is so prevalent~\cite{melzak,angle, angle2, redner, saving, ysm, money, Moukarzel, Villifane, rmp, ajp, Bouchaud, review, boghos, boghos2, review2}. One of the earliest models is  the yard-sale model~\cite{ysm} for which two agents are chosen at random, and a fraction $f$  of the poorer agent's wealth is randomly exchanged between the two agents. After these exchanges are repeated many times, one agent eventually  accrues all the wealth leading to wealth condensation. The  latter can  be prevented  by a bias of the exchange in favor of the poorer agent~\cite{Moukarzel,Villifane} or by some form of wealth redistribution. An example of the latter is    the growth, exchange, and distribution (GED) model~\cite{GED1, GED2} in which  the total wealth is increased  (to represent  the growth of the gross domestic product) by a fixed percentage, and  the added wealth is distributed  to all the agents so that the increased wealth of agent $i$ is   proportional to $w_i^{\lambda}$, where $w_i$ is the wealth of agent $i$, and $\lambda$ is the distribution parameter. After the distribution, the wealth of each agent  is rescaled so that the total wealth remains constant, and thus this model is an example fo a driven dissipative system. For $\lambda < 1$, there is no wealth condensation,  there is economic mobility, and the system can be described using equilibrium statistical mechanics even though multiplicative noise is present due to the exchange mechanism. For $\lambda \geq 1$, there is wealth condensation, no economic mobility, and the system is not in equilibrium. However, for all values of $\lambda$ the overall wealth distribution is not realistic, and the mechanism for the growth is treated as an external rather than as an internal mechanism.

The primary  goals of the following are to understand the role that multiplicative noise plays in a simple model of a driven dissipative system  and to discuss some simple  internal mechanisms for economic growth which  lead to  realistic wealth distributions.  For example, the cumulative wealth distribution $\Pi(w)$  of most nations~\cite{Pdist} shows  power law behavior for wealthy people, $\Pi(w) \sim w^{-\alpha}$, where $\alpha$ is typically between 1 and 3 and is known as the Pareto index~\cite{Pareto}. For poor people, $\Pi(w)$ exhibits   Boltzmann-like behavior,  $\log{\Pi(w)} \sim - \beta w$, where $\beta$ is a constant~\cite{Pdist}. We will find that our model leads to realistic behavior of the wealth distribution for both wealthy and poor agents, and realistic values of the Pareto index and the wealth and income Gini  coefficients~\cite{gini}.

The paper is organized as follows. In Sec.~\ref{IED} we introduce the model and discuss how it is simulated. In Sec.~\ref{results} we compare our results to  relevant  economic data, and in Sec.~\ref{sm} we discuss the statistical physics of the model. Concluding remarks are given in Sec.~\ref{conc}.

\section{The \gedi\ Model}\label{IED}

The GED model incorporates the wealth exchange mechanism of the yard-sale model, and a novel distribution mechanism of the increased wealth. However,  the growth of the wealth is assumed to be a constant which does not depend on the wealth of each agent. No mechanism for how this growth occurs is assumed. 

We generalize the GED model by replacing the externally imposed growth  with an internal investment mechanism that depends on the wealth of each agent~\cite{bm,scafetta}. In particular, after $N$ exchanges, the wealth of each agent is updated as follows
\be
w_i \to w_i + \FI (2r - 1 + g) w_i, \label{invest}
\ee
where $r$ is a uniform random number in the range $(0,1]$, the parameter $\FI$ is the fraction of the wealth that may be invested, and the parameter $g > 0$ represents the tendency of the wealth    to increase.  Because a new random number is  generated for each agent, some agents increase their wealth and some decrease their wealth as long as $g < 1$. If $g \ge 1$,  all investments lead to increases in wealth which is not realistic. 

This investment mechanism incorporates two realistic features. First, the greater the wealth of an agent,  the greater the average change in its wealth. Second, an agent's wealth may increase or decrease just as can occur in any investment. We shall find that the investment mechanism   is primarily responsible for the power law behavior of the wealth distribution for wealthy agents.

We  also give each agent a  small amount of wealth $\U$  after each unit of time.  This guaranteed income $\U$ crudely models a universal basic income or the fact that some of the wealth of all individuals is the same if we assume that public goods such as streets, police and fire departments, and other government services  contribute to each agent's wealth.

In addition to the exchange and investment mechanisms that we have described, we  incorporate the same wealth distribution mechanism used in the GED model.

In summary, the simulation of the  \gedi\ model (the generalized GED model with  investment) proceeds as follows.
\begin{enumerate}
\setcounter{enumi}{-1}

\item Assign the initial wealth of each agent to be  $w_i(t = 0)= 1$, so that the total wealth is $N$.

\item {\it Exchange wealth}. Choose agents $i$ and $j$ at random regardless of their
wealth and determine the amount $f\min[w_i(t), w_j(t)]$ to be exchanged.  Choose at random which agent gains and which agent loses~\cite{angle}.

\item {\it Investment and growth}. After $N$ exchanges, update the wealth of each agent according to Eq.~\eqref{invest} and determine $\mubar(t)$, the mean change in the wealth per agent at time $t$.

\item {\it The distribution mechanism}. If $\mubar(t) > 0$, assign the additional wealth due to growth to the agents
as~\cite{GED1}
\be
\label{distributionofgrowth}
\Delta w_{i}(t) = \mubar(t) W(t) \dfrac{w^{\lam}_{i}(t)}{\sum_{i=1}^{N} w^{\lam}_{i}(t)},
\ee
where $W(t)$ is the total wealth at time $t$ and $\mubar(t)$ is the mean growth rate at time $t$ averaged over all agents. If $\mubar(t) \leq 0$, there is no distribution and thus this step is skipped. 

\item {\it Guaranteed income}. Add $\U$ to the wealth of each agent after $N$ exchanges. 

\item  Rescale $w_i(t)$ so that $\sum_i w_i(t) = N$. 

\item Set $t = t + 1$. One unit of time corresponds to $N$ exchanges.

\item Repeat steps 1--6 until a steady state wealth distribution
is attained and then determine the average values of the desired quantities of interest.

\end{enumerate}
The wealth is scaled at the end of each unit of time  so that the total wealth remains constant. The rescaling crudely models the effect of inflation so that we   measure wealth in constant dollars. 

Note that the (average) increase in wealth due to investment is assigned  according to Eq.~\eqref{invest} and then the same amount is distributed according to Eq.~\eqref{distributionofgrowth}.  This distribution models the fact that investment does not just increase the wealth of  individual investors, but also  leads to gains for the economy as a whole. We discuss in Sec.~\ref{sec:Unonzero} the effects of distributing only a fraction of the increase in wealth. As in the GED model, this distribution   is equivalent to first order in $\mu$ to distributing  the revenue from a flat tax at a fixed rate $\mu$. In Ref.~\onlinecite{boghos} the revenue from a  tax at a fixed rate is distributed equally to all agents, which is equivalent to setting $\lambda = 0$ in the \gedi\  and the GED model. 

\section{Results}\label{results}
 
\begin{figure}[h]
\includegraphics[scale=0.5]{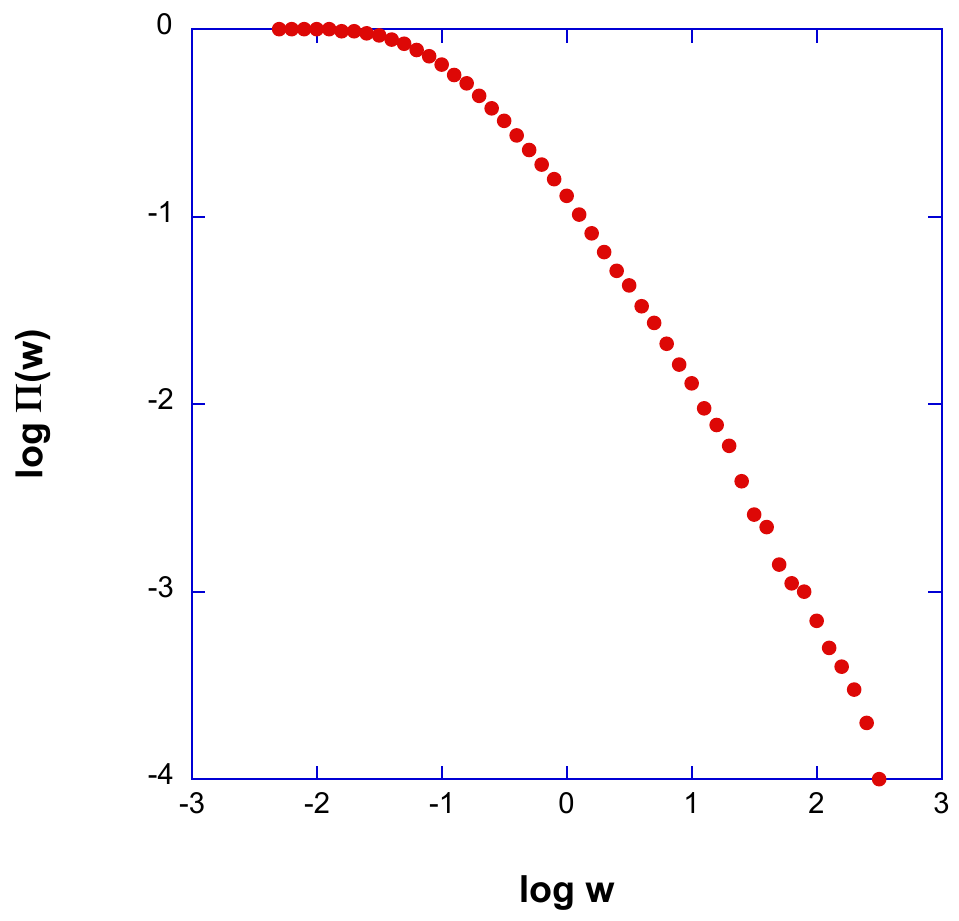}
\vspace{-0.2in}
\caption{The cumulative wealth distribution $\Pi(w)$ for the parameters  $f = 0.1$, $\lambda = 0.9$, $\FI= 0.8$, $g = 0.1$, and $\U = 0.01$,    with $N = 10^4$ at $t=5 \times 10^4$  after an equilibration time of $2 \times 10^4$. Three regions can be identified: a low wealth region that  shows    Boltzmann-like behavior  as shown in Fig.~\ref{fig1}, a high $w$ region that exhibits  power law behavior as shown in Fig.~\ref{fig2}, and a transition region between these two behaviors.} 
\label{fig0}
\end{figure}

\begin{figure}[h]
\includegraphics[scale=0.5]{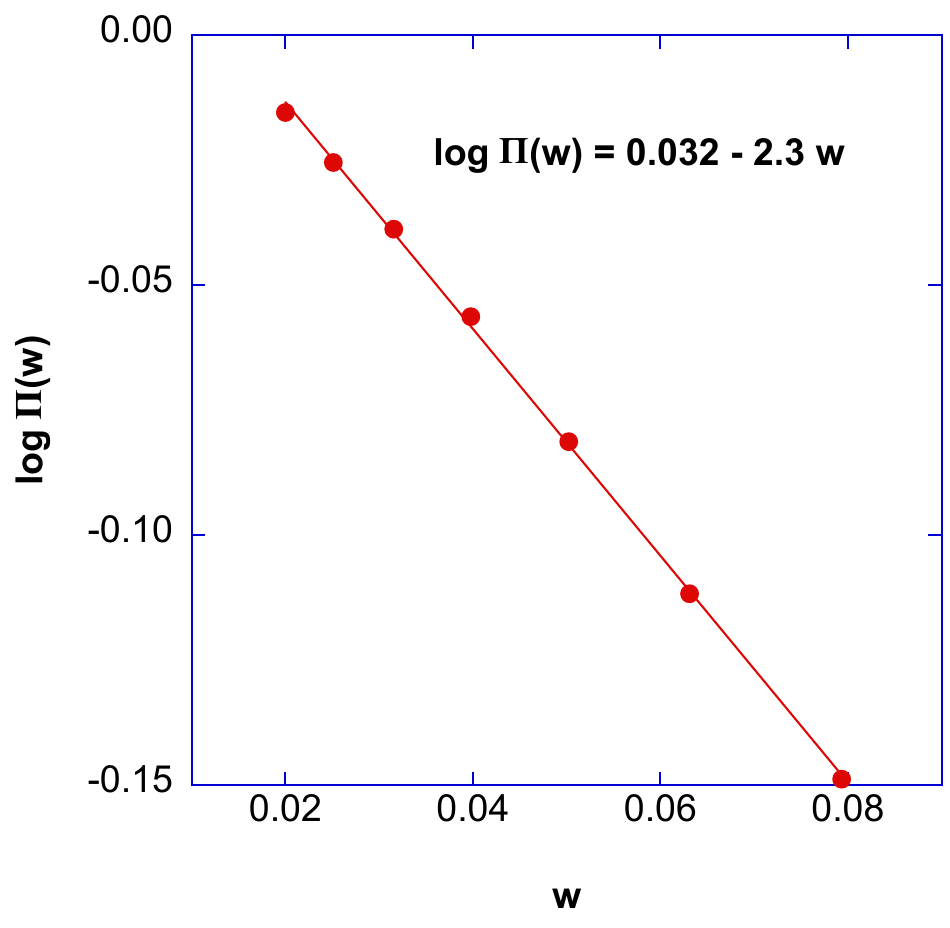}
\vspace{-0.2in}
\caption{The cumulative wealth distribution $\Pi(w)$ for poorer agnts using the same  parameters as in Fig.~\ref{fig0}. The straight line represents an exponential fit $\Pi(w) \sim e^{-\beta w}$ with $\beta \approx 2.3$  for  wealth $0.02 \leq w \leq 0.08$.}
\label{fig1}
\end{figure}

\begin{figure}[h]
\includegraphics[width=3.0 in]{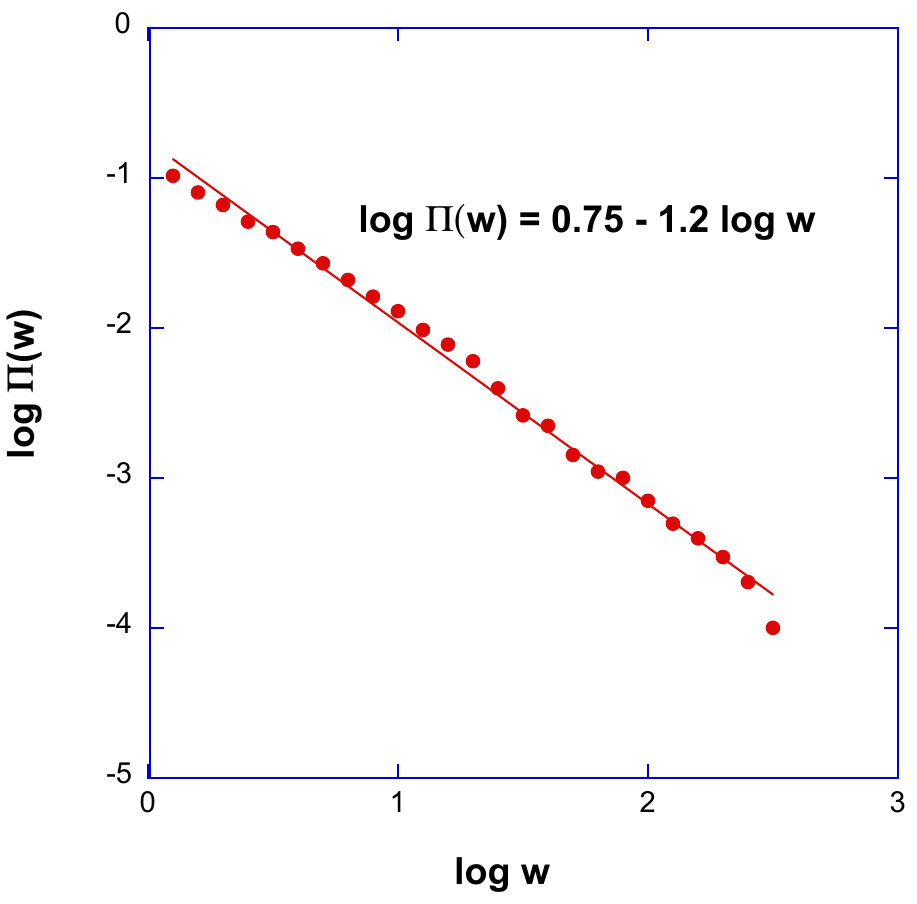}
\vspace{-0.2in}
\caption{The linear behavior of a log-log plot of $\Pi(w)$ for large wealth indicates a power law fit  with a Pareto index of  about $1.2$.}
\label{fig2}
\end{figure}

The cumulative wealth distribution $\Pi(w)$ is the fraction of the population with wealth greater than $w$ so that  $\Pi(0) = 1$. Our results for $\Pi(w)$ are shown in Figs.~\ref{fig0}--\ref{fig2} for $N = 10^4$ and the   parameters $f = 0.1$, $\lambda = 0.9$, $\FI = 0.8$, $g = 0.1$, and $\U = 0.01$. 
For low wealth $w$ we observe an exponential Boltzmann-like  distribution, as shown in the log-linear plot of $\Pi(w)$ in Fig.~\ref{fig1}. For  low values of $w$ the  wealth distribution is described by a power law with a Pareto index approximately equal to 1.2 for this choice of parameters as shown in Fig.~\ref{fig2}. Both of these behaviors are characteristic of real economic systems~\cite{Pdist} and are similar to what was found in Refs.~\cite{bm,scafetta}. The qualitative behavior of the cumulative probability distribution with  power law behavior for high wealth and Boltzmann-like behavior for low wealth  occurs for all values of $\lambda$ with $\U > 0$, and all values of $\lambda < 1$ for $\U = 0$.

As the investment fraction $\FI$ is reduced, the Pareto index increases. For  $\FI \lesssim 0.1$, the log-log plot of the cumulative wealth distribution at large wealth shows significant curvature. Thus, the investment mechanism, which generates  multiplicative noise, is the source of the power law distribution for large wealth. 

The investment mechanism leads to many agents obtaining large wealth, and to the power law behavior of $\Pi(w)$.  In contrast, no agent  in the  GED model  obtains a significant fraction of the total wealth for $\lambda < 1$~\cite{GED1}. For example, for $f = \mu = 0.1$ and $\lambda = 0.9$ with $N=10^4$, 
no agent has a wealth greater than 6. Even for $\lambda = 0.999$, no agent has a wealth greater than 200.  For the \gedi\ model 
with a comparable  average  growth rate of $\mubar = 0.08$ (using the same parameters  as in Figs.~\ref{fig0}--\ref{fig2}), we find that there are many agents with wealth greater than 1000~\cite{GED1}.

Common measures of wealth and income inequality are the Gini coefficients~\cite{gini}. Perfect equality corresponds to a Gini coefficient of zero and perfect inequality  corresponds to a Gini coefficient of one.  These  coefficients are frequently similar, but some countries such as those in Scandinavia have low income Gini coefficients, but large wealth Gini coefficients~\cite{sweden}.

To obtain realistic values of $\gi$, the income used to calculate the income Gini coefficient, $\gi$, includes    an added guaranteed income $U > 0$ as well as  the wealth gained by the  exchange mechanism. 
For the  parameters used in Fig.~\ref{fig0},     $G_w      \approx 0.84$, and the  $\gi   \approx 0.45$.    These numerical values of $G_w$ and $\gi$ are consistent with their values   for the United States  for which $G_w$  increased from 0.80 in 2008 to 0.85 in 2019 and has remained at 0.85  through 2021~\cite{giniW}. Similarly, $\gi$  increased from 0.43 to 0.49 from 1990 to 2020 with occasional fluctuations of magnitude 0.01, and then decreased to 0.47 in 2022~\cite{giniI}.

Another measure of the wealth distribution is the percentage of the wealth of  different sectors of the population.  For the parameters used in Fig.~\ref{fig0}, we find that the top 10\% of the population has 79\% of the wealth, the next 40\% has 17\% of the wealth, and the bottom 50\%  has 4\% of the total wealth, values which are  comparable to those found for the  United States in 2022 for which the percentages were 69\%, 8\%, and 3\%, respectively~\cite{aspen}.

\begin{figure}[h]
\includegraphics[width=3.0 in]{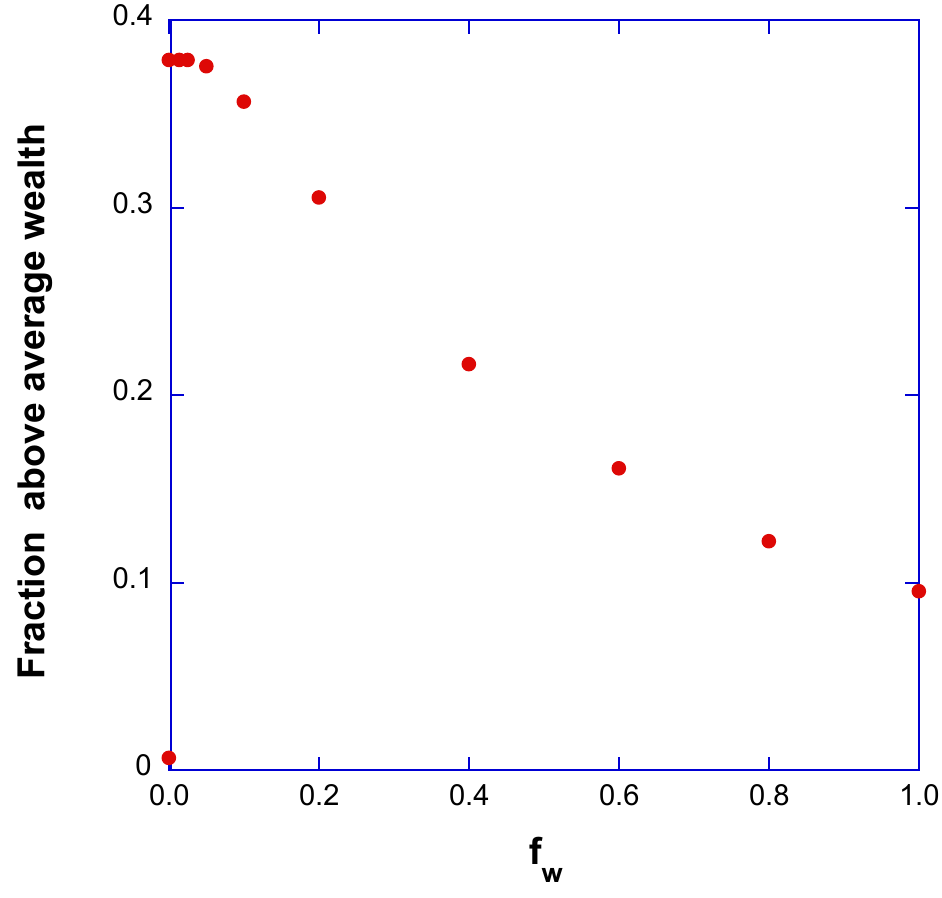}
\vspace{-0.1in}
\caption{The fraction of  agents with wealth greater than the average wealth of 1.0 as a function of the investment fraction $\FI$ for $\lambda = 0.9$. After the initial jump at $\FI = 0$, this fraction is a decreasing function of $\FI$. }
\label{above}
\end{figure}

Another measure of wealth inequality is the fraction of the population with wealth above the average wealth. 
This fraction has an interesting dependence on the investment fraction $\FI$. If  $\FI = 0$,  the \gedi\ model  reduces to the original yard-sale model plus the added income $\U$. The result is that eventually one agent gains almost all of the wealth and the other agents  have wealth  $\approx \U$. For $\FI = 0^+$ the fraction of agents with wealth above the average   is $\approx 0.38$, a value that increases  with $\U$, but is always less than 0.5. The fraction then slowly decreases as $\FI$ is increased (see Fig.~\ref{above}). The jump in $\FI$ from near 0 to 0.38 can be explained  as follows. Any investment leads to agents gaining and losing wealth, and prevents one agent from gaining all the wealth,  because richer agents can lose wealth through investment and overall gains in investment are distributed to  other agents. For small investment fractions, a significant fraction of the population  has a wealth slightly above average. As the investment fraction increases, the wealthiest agents benefit more, but the fraction of agents with wealth above the average declines. 

The value of the Pareto index for high wealth, the value of $\beta$ in the exponential distribution $\Pi (w) \sim e^{-\beta w}$ for low wealth,  
the range of the transition region between  Boltzmann-like behavior and power law behavior, and the values of the Gini coefficients all depend on the values of the various parameters (see Table~\ref{table1}). These quantities are not independent because increased inequality correlates with a greater Gini wealth coefficient, smaller Pareto index, larger $\beta$, and a smaller width of the transition region. For example, as $G_w$ increases, the Pareto index decreases, because a smaller Pareto index indicates that it is more probable to have agents with very large wealth leading to greater inequality.  Our numerical results  are independent of  $N$ for $N \gtrsim 1000$. 

The qualitative dependencies of various measures of the wealth distribution are summarized in  Table~\ref{table1}. For example,  $G_w$ increases with $\lambda$, because increasing $\lambda$ causes wealthier agents to receive more wealth from the distribution mechanism. As  the fraction of wealth invested $\FI$  increases, $G_w$  increases, because  wealthier agents   gain more by investment on the average.  As $g$ is increased,  there is more growth due to   investment benefiting all agents, and hence $G_w$ decreases.  
For  increases in $\FI$ or $\lam$,  the Gini income coefficient $\gi$ decreases, because as wealthy agents create more wealth, some of this  wealth  trickles down as income for poorer agents. Both Gini coefficients decrease as the guaranteed income $\U$ is increased, because  this income is given to all agents  and hence tends to equalize the income and then the wealth of the agents. 

\begin{table}[h!]
\centering
\begin{tabular}{|l|c|c|}
\hline
Parameter &  $G_w$  & $\gi$ \\
\hline
Distribution parameter $\lambda$ & increases &  decreases \\
Investment fraction $\FI$ & increases &   decreases \\
Growth parameter $g$  & decreases  & increases \\
Guaranteed income $\U$  & decreases & decreases \\
\hline
\end{tabular}
\caption{The qualitative dependencies of the  wealth and income Gini coefficients on  increases of the various parameters of the \gedi\ model. Other measures of inequality are correlated with the Gini coefficients as discussed in the text.  }  
\label{table1}
\end{table}

\begin{figure}[h]
\includegraphics[width=3.0 in]{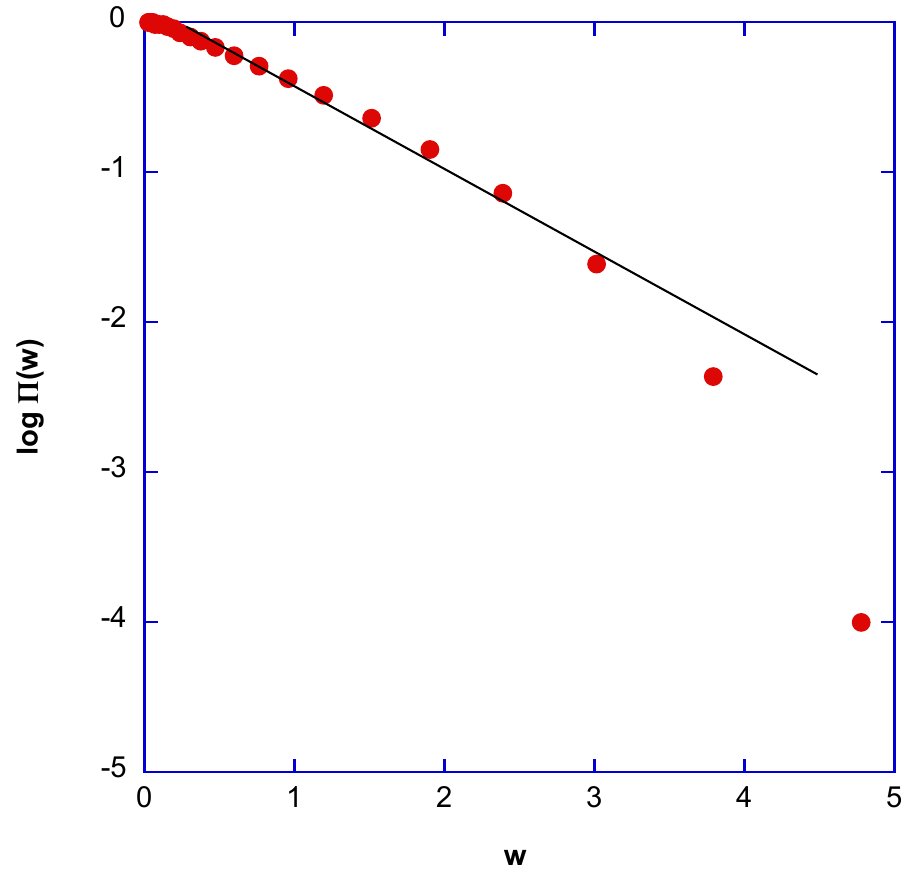}
\vspace{-0.2in}
\caption{The $w$-dependence of the cumulative wealth distribution $\log{\Pi(w)}$   for the GED model with the parameters  $f = 0.1$, $\lambda = 0.9$,  $\mu = 0.1$, and $N = 10^4$. The linear dependence for $ w \lesssim 3$ indicates a Boltzmann-like distribution similar to the behavior shown in Fig.~\ref{fig1} of the \gedi\ model for  $w\lesssim 0.08$, but there is no indication of the power law behavior for high wealth found in the \gedi\ model.}   
\label{GEDdist}
\end{figure}

The  three regions in the wealth cumulative probability distribution -- a small $w$ Boltzmann-like region, a large $w$ power law  region, and a cross-over region between these two behaviors can be interpreted as representing the lower, upper, and middle classes, respectively. The upper class achieves its wealth primarily through investment~\cite{bm,scafetta}. The exponential wealth distribution for the lower class is similar to what is found for the cumulative probability distribution  of the GED model  (see Fig.~\ref{GEDdist}). The middle class gains  from  investments, distribution of the growth from investments, and exchanges. The range of wealth of the middle class 
increases with increasing $\lambda$,  $\FI$, and $g$  because the  mechanisms associated with these parameters   increase the overall wealth in the society which benefits not just the wealthy but also the middle class. Increasing $\FI$ and $g$ directly increases the amount invested according to Eq.~(\ref{invest}), which in turn increases the total wealth in the system. The effect of increasing $\lambda$ is less direct  because    it puts more wealth in the hands of wealthier agents who then can invest more. As $g \to 0$, the growth approaches zero, and the middle class shrinks as more agents follow  Boltzmann-like behavior and fewer agents have large wealth. As $\U$ increases, more agents maintain their wealth through this added wealth. The effect is to shrink the middle class, but also to reduce inequality.

\begin{figure}[h!]
\includegraphics[width=3.0 in]{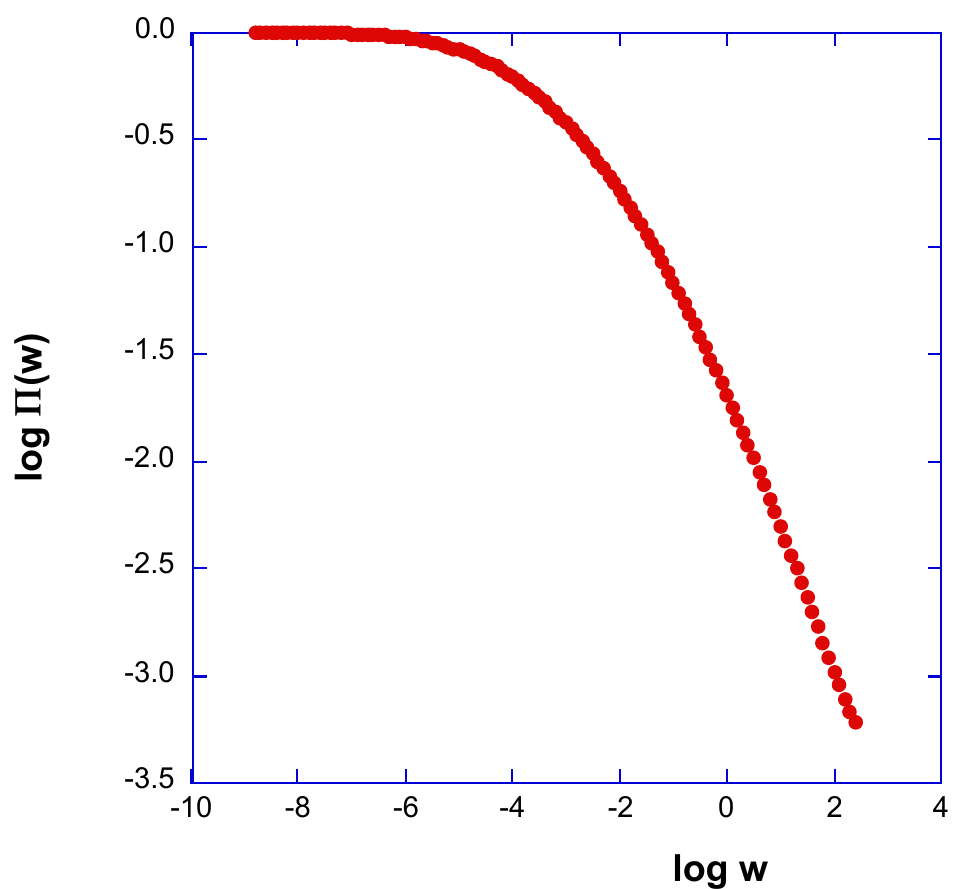}
\vspace{-0.1in}
\caption{The cumulative wealth distribution $\Pi(w)$ for the \gedi\ model with the same parameters as Fig.~\ref{fig0} except that $\U = 0$. At $\log{w} = 0$ ($w = 1$), $\log{\Pi(w)} \approx -1.7$ or $ \Pi(w) \approx 0.02$, which means 98\% of the agents have wealth less than $w=1$. Note the difference in the horizontal scales of Figs.~\ref{fig0}  and this figure.}
\label{figU=0}
\end{figure}

The  wealth distribution for $\U = 0$ is shown in Fig.~\ref{figU=0}. There are two  noticeable differences from the distribution for $\U>0$. As expected, there are many more poor agents. From Fig.~\ref{fig0} for $U=0.01$, we see that approximately 90\% of the agents have wealth less than the average wealth $w=1$; in contrast, for $\U= 0$ this percentage rises to 98\%. In addition, the Pareto index is about $0.65$, which is much smaller than the value $1.2$ found for $\U = 0.01$. Both features are indications of increased wealth inequality for $\U = 0$.  

\section{Statistical Physics of the \gedi\ Model}\label{sm}

\subsection{\label{sec:interest}Quantities of interest}

We first define  several quantities of interest  so that we can contrast   their behavior in the \gedi\ and GED models. A key input parameter of the GED model is the growth rate  $\mu$. In the \gedi\ model the average growth rate is determined by the input parameters $\FI$ and $g$. If we sum over the investment of all the agents using  Eq.~(\ref{invest}), we find that the mean growth rate is
\be
\mubar = g \FI,  \label{mu}
\ee
because the sum over the random numbers $(2r-1)$ is approximately zero.  
Equation~\eqref{mu} was  confirmed numerically. 

The order parameter is taken to be~\cite{GED1}
\begin{equation} 
\phi = \frac{N- w_{\max}}{N},
\end{equation}
where $w_{\max}$ is the wealth of the richest agent. As $N \to \infty$,  $\phi \to 1$ when there is no wealth condensation and $\phi \to 0$ when all the wealth condenses into one agent. The susceptibility per agent $\chi$ is defined as the variance of the wealth of each agent averaged over all agents~\cite{chidef}.

The energy is defined as~\cite{GED1, GED2}
\begin{equation} 
E = \sum_{i = 1}^N [1-w_i(t)]^2.
\end{equation} 
This definition minimizes the energy when the wealth is equally distributed, and behaves like the energy in thermal systems for the GED model. 

To determine if the GED and \gedi\ models are effectively ergodic, we consider the (rescaled) wealth metric as~\cite{tm}
\be
\label{eq:metric}
\Omega(t) = \frac{1}{N}\sum_{i=1}^{N}\big [ {\overline w}_{i}(t) - \overline{w}(t) \big]^{2},
\ee
where $\overline{w}_{i}(t)$ is the time averaged wealth of agent $i$ at time $t$,
\begin{align}
\overline{w}_{i}(t) &= \frac{1}{t}\!\int_{0}^{t} w_{i}(t')\,dt', \\
\noalign{\noindent and $\overline{w}(t)$ is the average over all agents,}
\overline{w}(t) &= \frac{1}{N} \sum_{i=1}^{N}{\overline w}_{i}(t) = 1.
\end{align}
If the system is effectively ergodic, $\Omega(t)\propto 1/t$~\cite{tm}. Effective ergodicity is a necessary, but not a sufficient condition for ergodicity. The    GED model  is effectively ergodic for $\lam < 1$, but is not ergodic for $\lam \geq 1$ for which  wealth condensation occurs~\cite{GED1}.

It is  difficult to determine whether  a system is in thermal equilibrium  or in a nonequilibrium steady state.  In both cases macroscopic quantities are independent of time. A  subtle measure of nonequilibrium behavior was  introduced by Zia and Schmittmann~\cite{zia}. The essence of the method is to define an analogy to the angular momentum $L$ for two quantities as
\begin{equation}
L = (A(t) - {\bar A})  (B(t+ \Delta t) - {\bar B}) -  (A(t+\Delta t) - {\bar  A}) (B(t) - {\bar B}),  \label{Ldef}
\end{equation}
where $A$ and $B$ are two measurable quantities. 
If there is a net ``circulation" indicated by a nonzero value of  $L$,  the system is   in a steady state rather than in equilibrium. A sensitive measure of the circulation is the asymmetry in $L$ given by~\cite{zia} 
\begin{equation} 
\Upsilon(L) = \frac{H(L) - H(-L)} {[H(L) + H(-L)]^{1/2}},  \label{asym}
\end{equation} 
where $H(L)$ is the histogram of $L$ values.
If the absolute values of $\Upsilon(L)$  are systematically greater than unity,  we conclude the system is in a steady state rather than in equilibrium. The use of Eq.~(\ref{asym})  can   only demonstrate that a system is not in equilibrium if asymmetry is found; however, if no asymmetry is found, we  can   conclude  only that the measurement is consistent with  the system being in equilibrium, because a lack of asymmetry may exist for some variables but not others.  For the GED model with $A$ equal to the Gini coefficient and $B$ equal to the order parameter, we find that the average value of $L$ is negligible, and thus there is no asymmetry. This lack of asymmetry is consistent with thermal equilibrium as concluded in Ref.~\cite{GED1} using a less stringent criterion.

We  discuss the behavior of the quantities introduced in this section for the \gedi\ model for $\U = 0$  in Sec.~\ref{sec:Uzero}  and  for $\U > 0$ in Sec.~\ref{sec:Unonzero}.

\subsection{\label{sec:Uzero}$\U = 0$}

For both the GED model and the \gedi\ model with $U = 0$,   $\phi = 1$ for $\lambda < 1$ and $\phi = 0$ for $\lambda \geq 1$ in the limit $N \to \infty$, indicating there is a phase transition at $\lam=1$.  One difference is that the limiting behavior $\phi \to 1$ as $N \to \infty$  for $\lambda < 1$  occurs much more slowly with $N$ for the \gedi\ model.
 
 For $\lam \geq 1$,  $\chi = 0$ for both the GED and the \gedi\ model with $U=0$,  and wealth condensation occurs.  The susceptibility $\chi$ of the GED model is characterized by $\chi \sim(1-~\lambda)^{-\gamma}$, with $\gamma = 1$, both at fixed $N$ and fixed Ginzburg parameter $G$ [see  Eq.~\eqref{G0}].

We now determine if the behavior of the \gedi\ model   for $\U = 0$ is similar to     the GED model near $\lam = 1$. In Refs.~\cite{GED1, GED2}   it was necessary  to keep   $G$ constant as $\lam \to 1$ to obtain a  thermodynamic description consistent with mean-field theory. In particular,  this condition was necessary to obtain the mean-field value of the critical exponent of the specific heat.   The condition of keeping   $G$ constant as $\lam \to 1$  allows us to reduce the effects of  the multiplicative noise due to the exchange mechanism. In the \gedi\ model  there is additional multiplicative noise due to the investment mechanism. This noise cannot be reduced as $N \to \infty$ because  investment leads to  noise in the growth  of the total wealth, which in turn affects the wealth of all the agents due to the distribution mechanism. Nevertheless, we will explore what happens for the same conditions used for the GED model. 

We adopt the same definition of the Ginzburg parameter  used in Refs.~\cite{GED1, GED2} with $\mu$ replaced by $\mubar$:
\begin{equation}
G = \frac{N \mubar_0 (1- \lambda)}{f_0^2}, \label{G0}
\end{equation}
where $f_0$ and $\mubar_0$ are the values of $f$ and $\mubar$ at $N=N_0$. The  growth parameter $\mubar$ and the exchange parameter $f$ used in the simulations is scaled as $\mubar =  \mubar_0N_0/N$ and $f = f_0N_0/N$, respectively. Thus,
To obtain a consistent description of the energy and the specific field in the GED model with mean-field theory, it is necessary to keep $G$ fixed  as $\lam$ is varied and hence $N$ must increase  as $\epsilon \equiv 1-\lambda$ decreases.  In addition, to  minimize  the  multiplicative noise associated with  the exchanges of wealth, we require that~\cite{GED2}
\begin{equation}
M \equiv \frac{ \sqrt{N}\mu_0 (1- \lambda)}{f_0}  \gg 1.  \label{M}
\end{equation}

\begin{figure}[tbp]
\includegraphics[scale=0.4]{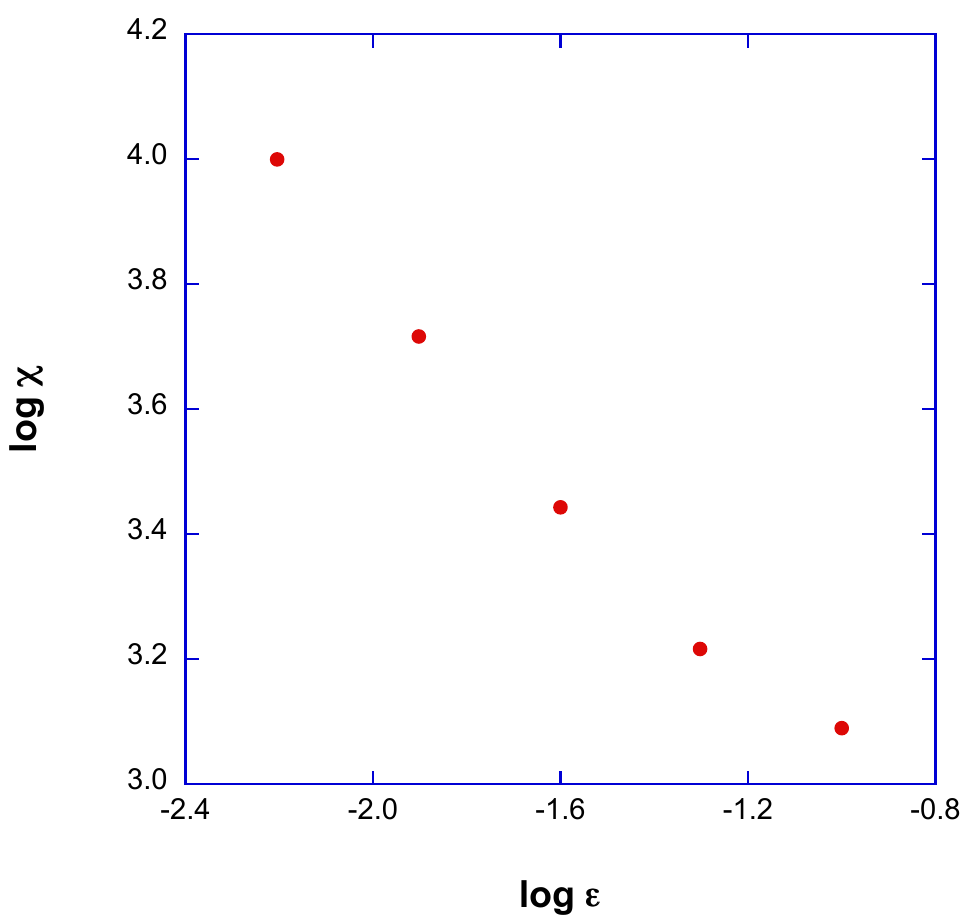}
\includegraphics[scale=0.4]{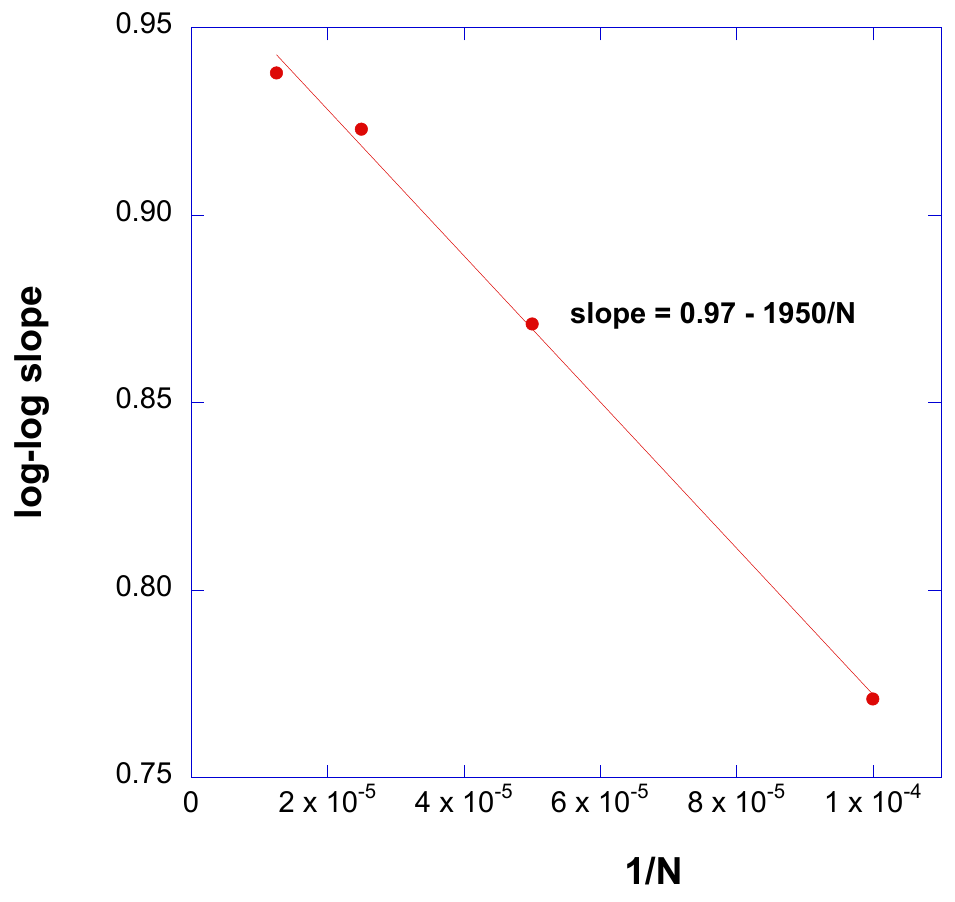}
\vspace{-0.5cm}
\caption{\label{susc}(a) The $\epsilon=1-\lam$ dependence of the susceptibility $\chi$  for fixed Ginzburg parameter $G = 10^6$ of the \gedi\ model with $\U = 0$.  The curvature of the log-log plot decreases as the critical point is approached and $N$ is increased.   (b) The slope of the log-log plot in (a) as a function of $1/N$, where  $N$ is the smallest number of agents used to compute the slope from the data points in (a). Thus, for $1/N =  10^{-4}$ all five points are used, and for $1/N = 1.25 \times 10^{-5}$ only the last two data points are used to determine the slope. 
A linear fit  yields $\gamma \approx 0.97$ in the limit $N \to \infty$.}
\end{figure}

For the data points shown in Fig.~\ref{susc}, $f_0 = 0.01$, ${\FI}_{,0} = \mubar_0 g = 1.0$,  $g= 0.1$, and $N_0 = 10^4$, so that $\mubar_0 = 0.1$.  We choose $G = 10^6$ so that  $25 \leq M \leq  100$, and $10^4 \leq N \leq  1.6 \times 10^5$. The  results for $\log{\chi}$ versus  $\log{\epsilon}$ in Fig.~\ref{susc}(a) for the \gedi\ model  show some curvature but the slope in the limit as $N\to \infty$ shown in Fig.~\ref{susc}(b) approaches  $\gamma = 1$ found for the GED model~\cite{GED2}.

In the GED model $\gamma$ can also be found from the divergence of the susceptibility at fixed $N$. In contrast,  for the \gedi\ model the log-log plot of $\chi$  versus $\epsilon=1-\lam$ for  $N=10^4$, $f=0.1$, $g=0.1$ and ${\FI} = 0.8$ exhibits significant curvature for $\epsilon \ll 1$.  This curvature may be due to the much larger fluctuations of the wealth in the \gedi\ model due to the multiplicative noise in the investment mechanism. In the \gedi\ model the susceptibility exponent converges to $\gamma = 1$ more quickly if  $G$ is held fixed. 

\begin{figure}[tbp]
\includegraphics[scale=0.5]{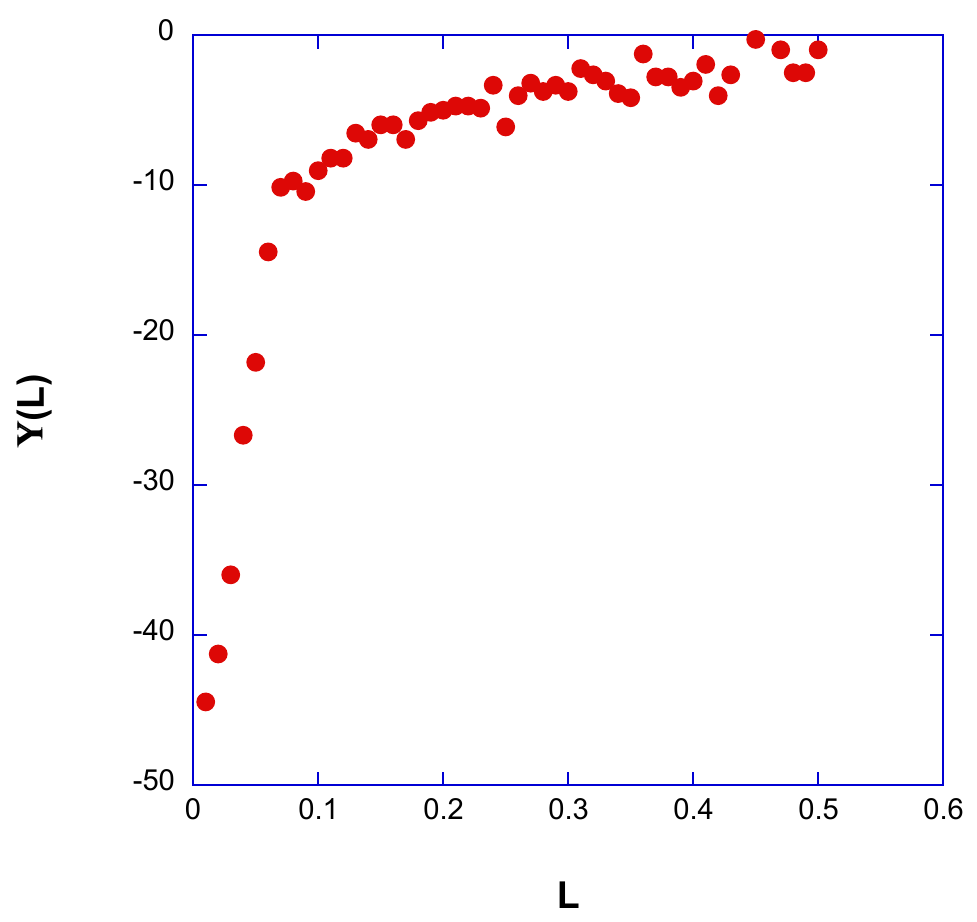}
\vspace{-0.5cm}
\caption{The asymmetry $\Upsilon(L)$  for $f = 0.1$, $\lambda = 0.9$,  $\FI= 0.8$, $g = 0.1$, and $\U = 0.0$ with  $N = 2500$. 
Because
$|\Upsilon(L)| > 1$ for a range of $L$ values, we conclude that the \gedi\ model is not in thermal equilibrium.} 
\label{asymU0}
\end{figure}

The wealth metric $\Omega(t)$ of the \gedi\ model for $\U = 0$ and $\lambda < 1$ shows the same behavior as in the GED model so both models are effectively ergodic.  However, if we determine the asymmetry of the probability angular momentum  using Eq.~(\ref{asym})   with the quantities $\mubar(t)$ and $\phi(t)$, we find evidence for steady state rather than equilibrium behavior as shown in Fig.~\ref{asymU0}, which is different from what was found for the GED model. This asymmetry is   independent of the number of agents $N$.  

\subsection{\label{sec:Unonzero} $\U > 0$}

For $\U > 0$ there is no  transition at $\lambda = 1$, and  $\phi > 0$ for all $\lambda$, because all agents regularly receive some added wealth,  and thus the wealthiest agent  never obtains all the wealth.  We can think of $\U > 0$ as analogous to  a magnetic field in the Ising model. For example, for  $\lambda = 1.1$ and all other parameters the same as in Fig.~\ref{fig0}, the top 10\% of the agents have 88\% of the wealth and the bottom 50\% have 2\% of the wealth. In contrast, if $\U=0$ and  $\lambda > 1$, one agent obtains all the wealth as in the   yard-sale and GED  models.  For all values of $\lambda$, the susceptibility  increases with $N$ as   $\chi \sim N^{a}$, where $0 < a \le 1$ depends on both $\lambda$ and $\U$, with $a \to 1$ as $\lambda \to \infty$.

Because the wealth distribution in the \gedi\   model is characterized by a  power law for large wealth (see Fig.~\ref{fig2}),   there are many agents with large  wealth. The large fluctuations of the wealth  of these agents    leads to the power law dependence of the susceptibility on $N$.  
The divergence of the susceptibility $\chi$ is analogous to two phase coexistence, 
where the compressibility in the  two phase coexistence of a liquid and gas vanishes   so that its inverse diverges.  The analog of the liquid phase is the large number of poor agents and the analog of the gas phase is the small number of wealthy agents. As $\lambda$ increases, the number of agents with wealth significantly greater than the mean $w= 1$ decreases, just as the density of gas molecules decreases  as the volume increases for two phase coexistence at fixed pressure for fluids.

For $\lambda = 1$, there is a transition as  $\U \to 0^+$  such that the susceptibility jumps from infinity to zero. If we   fit the $\U$-dependence of $\phi$ to $\phi \sim \U^b$  at $\lambda = 1$, we find $b \approx 0.03$, which suggests that the dependence  might be logarithmic. 

\begin{figure}[tbp]
\includegraphics[scale=0.5]{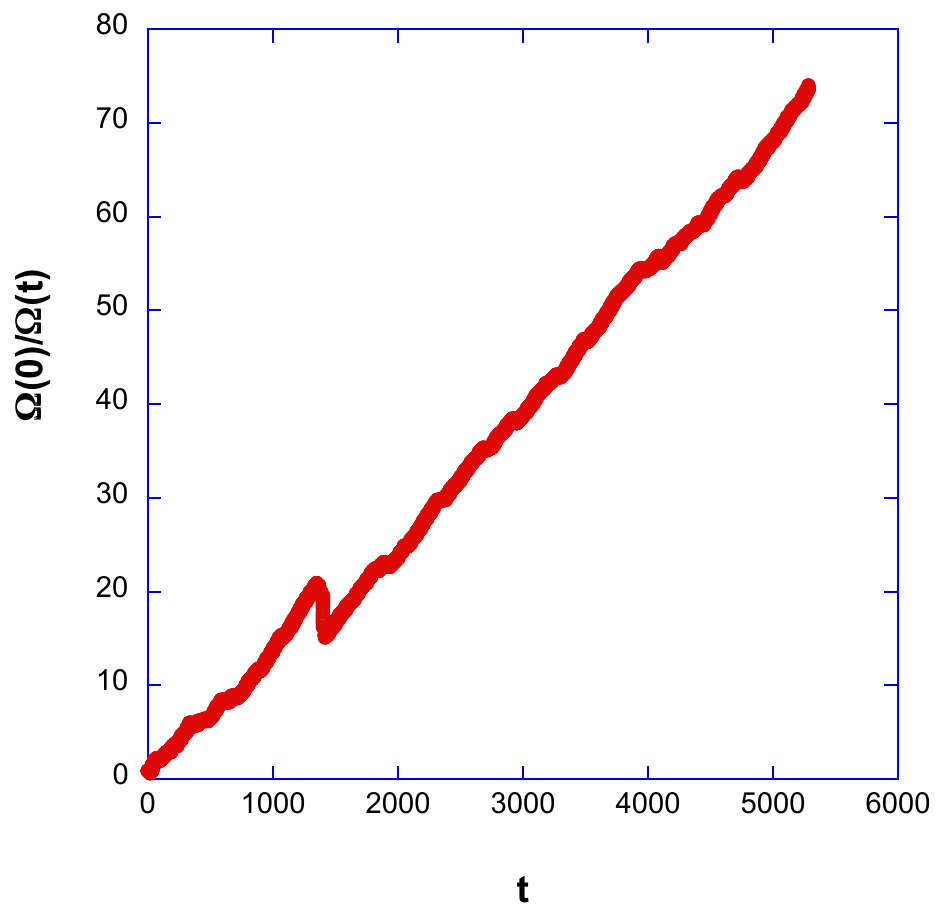}
\vspace{-0.5cm}
\caption{\label{metric} The linear time-dependence of the inverse  normalized wealth metric $\Omega(0)/\Omega(t)$ for the \gedi\ model with $U>0$ indicates that the system is in a steady state. The    parameters are the same as in Fig.~\ref{fig0}. Data was taken after  an equilibration time of about $2 \times 10^4$. The large deviation from linear behavior near $t=1300$ is due to a large fluctuation in the growth rate shown in Fig~\ref{growth}. }
\end{figure}

 For $\U > 0$ the system is effectively ergodic for all $\lambda$. An example is shown by the linear behavior of $\Omega(0)/\Omega(t)$  in Fig.~\ref{metric}.
This behavior is consistent with  the observation that the identity of the wealthiest agent fluctuates with time for all $\lambda$, which in turn indicates that there is economic mobility.  In contrast, the identity of the wealthiest agent in the GED model and the \gedi\ model with $\U=0$ does not fluctuate with time for $\lam \geq 1$.

\begin{figure}[tbp]
\includegraphics[scale=0.5]{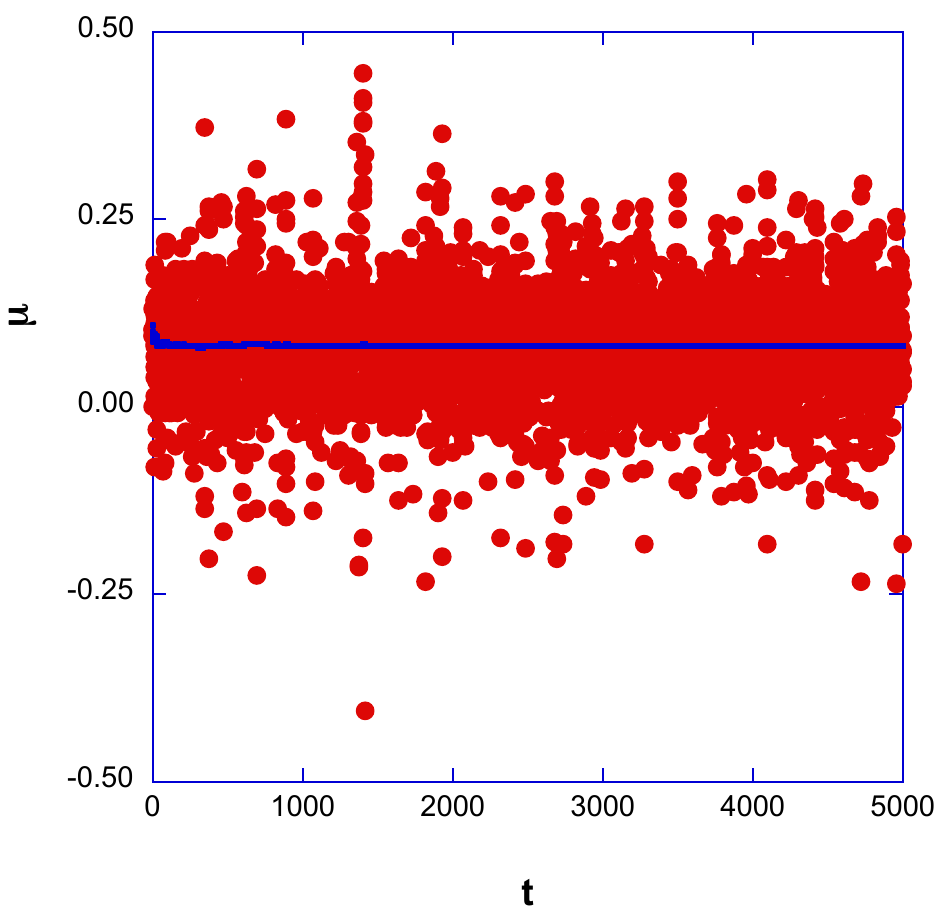}
\vspace{-0.5cm}
\caption{\label{growth} The time-dependence of the  growth rate $\mubar$ averaged over all agents for the same parameters as in Fig.~\ref{fig0}. The mean growth rate is about $0.08$, as  shown by the horizontal blue line. The large deviation from the mean  near $t=1300$ occurs at the same time as the large deviation from linear behavior in the inverse wealth metric shown in Fig.~\ref{metric}. }
\end{figure}

In Fig.~\ref{growth} we show $\mubar(t)$, the mean growth rate per agent. The occasional  large values of  $\mubar(t)$ correspond to the sharp deviations   from linear behavior of $\Omega(0)/\Omega(t)$ as shown in Fig.~\ref{metric}.  Similar behavior was found for the OFC model of earthquakes~\cite{OFC}. Note that  the growth rate is sometimes negative, indicating an overall loss due to investment at that time. 

\begin{figure}[tbp]
\includegraphics[scale=0.5]{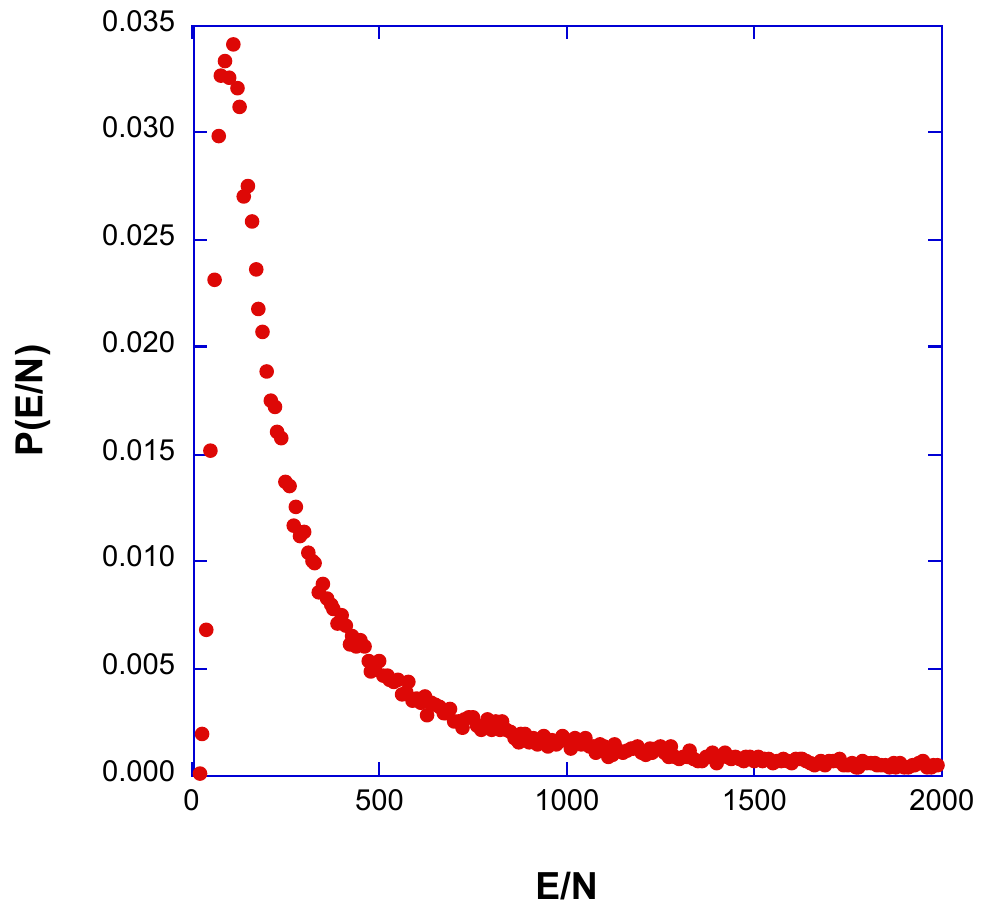}
\vspace{-0.5cm}
\caption{The energy probability density of the \gedi\ model for the same parameters as in Fig.~\ref{fig0}. } 
\label{EProb}
\end{figure}

The distribution of the energy $E$ in the GED model follows a Gaussian distribution for $\lambda < 1$ if $G$ and $M$ are sufficiently large. Also, the ratio of the energy distributions at two values of $f$ is consistent with a  Boltzmann distribution, suggesting that the system is in equilibrium similar to thermal equilibrium systems and consistent with the symmetry of $\Upsilon(L)$  discussed earlier. 
In contrast,  $E$ does not satisfy a Gaussian distribution for the \gedi\ model for $\U \geq 0$, but instead has a long tail at large $E$ similar to a log-normal distribution as shown in Fig.~\ref{EProb}. A log-normal distribution is expected when there is multiplicative noise~\cite{ole-bill}, which in the \gedi\ model is due in part to the   investment mechanism. See Ref.~\cite{ole-bill} for a discussion of log-normal distributions due to  multiplicative noise. 

\begin{figure}[tbp]
\includegraphics[scale=0.5]{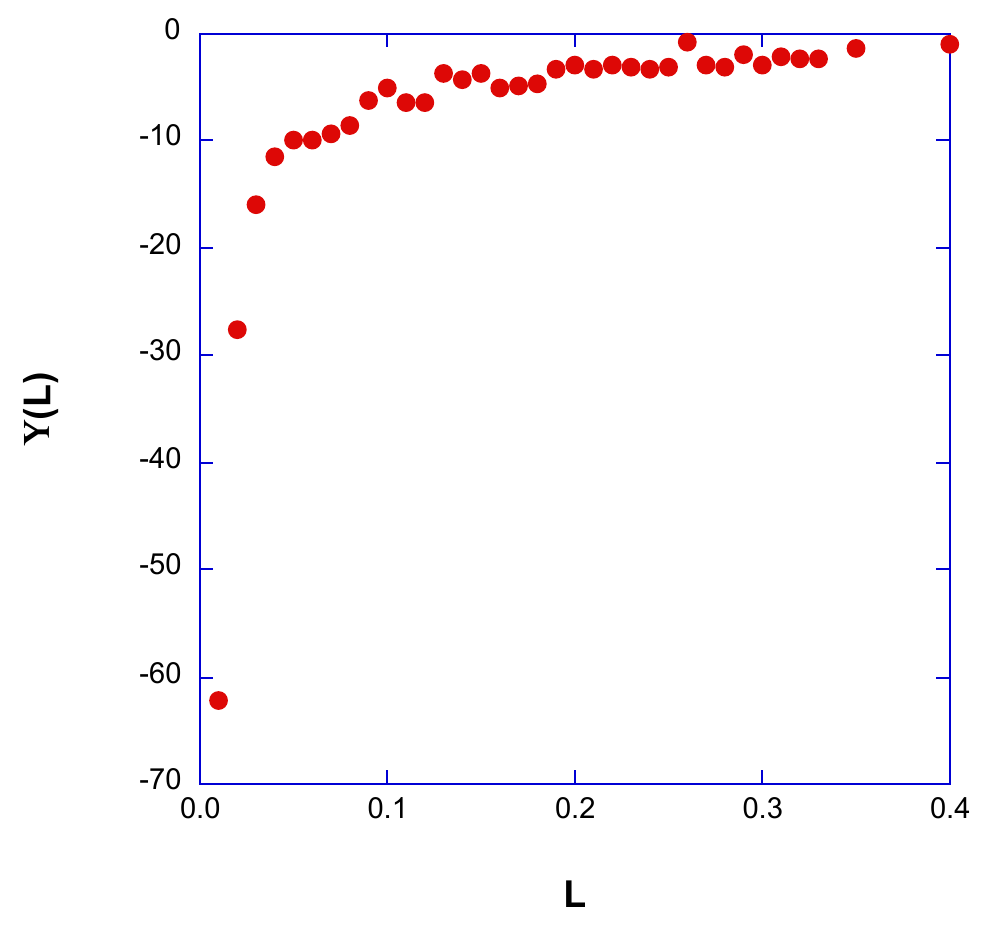}
\vspace{-0.5cm}
\caption{ Asymmetry $\Upsilon(L)$ for $N = 2500$,  $f = 0.1$, $\lambda = 0.9$, $\FI= 0.8$, $g = 0.1$, and $\U = 0.01$. Because
$|\Upsilon(L)| > 1$ for a range of $L$ values,  the system is not in thermal equilibrium.} 
\label{asymU}
\end{figure}

As for $U=0$, we calculated the asymmetry of the probability angular momentum  using Eq.~(\ref{asym})   with the quantities $\mubar(t)$ and $\phi(t)$, and find evidence for steady state rather than equilibrium behavior as shown in Fig.~\ref{asymU}. 

If we distribute only a fraction of the investment gain rather than all of it, we find small  changes in the Gini coefficients, the Pareto index, and the order parameter. For example, using the same parameters as in Fig.~\ref{fig0} and a fraction of $0.3$,  $G_{\rm in}$, the Pareto index, and $\phi$  decrease by about $10\%$, and   $G_w$ increases by about  $5\%$. However, the susceptibility increased  from $190$ when   all the investment gain  was distributed to $583$ for the $0.3$ fraction.  This large increase in the susceptibility is due to the fact that the distribution mechanism tends to reduce fluctuations, which is not reduced as much if only a fraction of the investment gain is distributed.

\section{Discussion} \label{conc}

Agent based models provide alternatives to the  approaches frequently used to study economic systems. 
The tools of statistical physics can  help us better analyze economic systems and determine whether the system is in a steady state or in equilibrium.  From a public policy point of view we can see how changing various parameters can lead to more or less wealth inequality. 

We have discussed the behavior of a  model of the accumulation and distribution of wealth. The model includes internal mechanisms for investment, exchange,  distribution, and a guaranteed income and yields  realistic wealth distributions. The resultant wealth distributions show a Pareto power law for the wealthy and an exponential distribution for the poor and are similar to what are seen in real economies. 
The investment mechanism is the primary source of the large wealth of a small fraction of the population, and the added income represents    that there are  some public goods that  are shared by all agents. The guaranteed income leads to realistic income Gini coefficients.

From the perspective of statistical physics the GEDI model is a relatively simple example of a driven dissipative system. This class of systems  has many examples,  including  biological and economic systems as well as earthquake faults.

The  \gedi\ model shows steady state behavior with occasional  large deviations from the average growth rate. There are two sources of multiplicative noise, that due to exchange which can be minimized in the limit $N \to \infty$ and that due to investment which cannot be minimized. The latter multiplicative noise is  responsible  for the nonequilibrium behavior of the \gedi\ model. The occasional large fluctuations  of the wealth in the \gedi\ model   can be seen in the time dependence of the wealth metric and in the growth rate. Similar behavior is seen in economic systems where there is short time steady state behavior, but over long times there are occasional large scale  changes in the economy. A difficult next step would be to see if there is any behavior in the \gedi\ model that is a precursor to such large deviations.

\begin{acknowledgments}
We would like to thank   Amy Galick, Bill Gibson, George Tuthill, and Royce Zia for useful discussions.
\end{acknowledgments}


\begin{thebibliography}{88}

\bibitem{melzak} Z. A. Melzak, {\it Mathematical Ideas, Modeling and Applications, Volume II of Companion to Concrete Mathematics} (Wiley, New York, 1976), p. 279.

\bibitem{angle}J. Angle, ``The surplus theory of social stratification and the size distribution of personal wealth,'' Social Forces {\bf 65}, 293 (1986).

\bibitem{angle2} J. Angle, ``Deriving the size distribution of personal wealth from the rich get richer, the poor get poorer,'' J. Math. Sociology {\bf 18}, 27 (1993).

\bibitem{redner} P. L. Krapivsky and S. Redner,``Wealth distributions in asset exchange models," Eur. Phys. J. B {\bf 2}, 267 (1998). 

\bibitem{saving} A. Chakraborti and B. K. Chakrabarti, ``Statistical mechanics of money: how saving propensity affects its distribution,'' Eur. Phys. J. B {\bf 17}, 167 (2000).

\bibitem{ysm}A. Chakraborti, ``Distributions of money in model markets of economy,'' Int. J. Mod. Phys. C {\bf 13}, 1315 (2002).

\bibitem{money}A. Dr\"agulescu and V. M. Yakovenko, ``Statistical mechanics of money,'' Eur. Phys. J. B {\bf 17}, 723 (2000).

\bibitem{Moukarzel}C. F. Moukarzel, S. Goncalves, J. R. Iglesias, M. Rodriguez-Achach, and R. Huerta-Quintanilla, ``Wealth condensation in a multiplicative random asset exchange model,'' Eur. Phys. J.-Spec. Top. {\bf 143}, 75 (2007).

\bibitem{Villifane} Gast\`on Villafa\~nea, Lautaro Giordanob, and Mar\`ia Fabiana Lagunac, ``Wealth inequality in agent-based economies:
The dominant role of social protection over growth,"  arXiv:2508.06666v1[physics.soc-ph] (2025). 

\bibitem{rmp} V. M. Yakovenko and J. B. Rosser Jr., ``Colloquium: Statistical mechanics of money, wealth, and income,'' Rev. Mod. Phys. {\bf 81}, 1703 (2009).

\bibitem{ajp}M. Patriarca and A. Chakraborti, ``Kinetic exchange models: From molecular physics to social science,'' Am. J. Phys. {\bf 81}, 618 (2013).

\bibitem{Bouchaud}J.-P. Bouchaud and M. M\'ezard, ``Wealth condensation in a simple model of economy,'' Physica A {\bf 282}, 536 (2000).

\bibitem{review}A. Chakraborti, I. M. Toke, M. Patriarca, and F. Abergel, ``Econophysics review: II. Agent-based models,'' Quant. Finance {\bf 11}, 1013 (2011).

\bibitem{boghos} B. M. Boghosian, ``Kinetics of wealth and the Pareto law,'' Phys. Rev. E {\bf 89}, 042804 (2014).

\bibitem{review2}Max Greenberg and H. Oliver Gao, ``Twenty-five years of random asset exchange modeling,'' Eur. Phys. J. B {\bf 97}, 69 (2024).

\bibitem{boghos2}B. M. Boghosian, ``Fokker-Planck description of wealth dynamics and the origin of Pareto's law,'' Int. J. Mod. Phys. C {\bf 25}, 1441008 (2014).

\bibitem{GED1} Kang K. L. Liu, N. Lubers, W. Klein, J. Tobochnik, B. M. Boghosian, and H. Gould, ``Simulation of a generalized asset exchange model with economic growth and wealth distribution," Phys. Rev. E {\bf 104}, 014150 (2021).

\bibitem{GED2}W. Klein, N Lubers, K. K. L. Liu, T. Khouw, and H. Gould, ``Mean-field theory of an asset exchange model with economic growth and wealth distribution," Phys. Rev. E {\bf 104}, 014151 (2021).

\bibitem{Pdist} Adrian Dr\v{a}gulescu and  Victor M. Yakovenko, ``Exponential and power-law probability distributions of wealth and income in the United Kingdom and the United States," Physica A {\bf 299}, 213 (2001). 

\bibitem{Pareto}V. Pareto, Cours d'Economie Politique, Lausanne (1897).

\bibitem{gini}C. Gini, "Concentration and dependency ratios," (1909). English translation in Rivista di Politica Economica, {\bf 87} 769 (1997);
C. Gini, ``Variabilit{\'a} e Mutuabilit{\' a}," Contributo allo Studio delle Distribuzioni e delle Relazioni Statistiche, Bologna: C. Cuppini (1912). An accessible introduction to the Gini coefficient can be found at \url{<https://en.wikipedia.org/wiki/
Gini_coefficient>}. 

\bibitem{bm} J. P. Bouchaud  and M. Mezard, ``Wealth condensation in a simple model of economy,"  Physica A {\bf 282}, 536 (2000).

\bibitem{scafetta} Nicola Scafetta, Sergio Picozzi and Bruce J. West, ``An out-of-equilibrium model of the distributions of wealth," Quantitative Finance {\bf 4}, 353 (2004); ``A trade-investment model for distribution of wealth," Physica D {\bf 193}, 338 (2004). 

\bibitem{sweden} \url{<https://en.wikipedia.org/wiki/Income_inequality_in_Sweden>}.

\bibitem{giniW} \url{<https://en.wikipedia.org/wiki/List_of_countries_by_wealth_inequality>}.

\bibitem{giniI} \url{<https://www.statista.com/statistics/219643/gini- coefficient-for-us-individuals-families-and-households/>}.

\bibitem{aspen}\url{<https://www.aspeninstitute.org/blog-posts/charts-that-explain-wealth-inequality-in-the-united-states/>}.

\bibitem{tm} D. Thirumalai and R. D. Mountain, ``Activated dynamics, loss of ergodicity, and transport in supercooled liquids,'' Phys. Rev. A {\bf 42}, 4574 (1990) and Phys. Rev. E {\bf 47}, 479 (1993).

\bibitem{chidef} In Refs.~\cite{GED1, GED2}  on the GED model the susceptibility is defined as $N$ times the average variance of the wealth of each agent. In this paper we define $\chi$ without the factor of $N$ so that a divergence of $\chi$ is the divergence due to the average variance and not due to the factor $N$. 

\bibitem{ole-bill} O. Peters and W. Klein, ``Ergodicity breaking in geometric Brownian motion,'' Phys. Rev. Lett. {\bf 110}, 100603 (2013).

\bibitem{zia}  R. K. P. Zia and B. Schmittmann, ``Probability currents as principal characteristics in the
statistical mechanics of non-equilibrium steady states," J. Stat. Mech. P07012 (2007);
R. K. P. Zia, ``Signals of detailed balance violation in nonequilibrium stationary states: Subtle, manifest, and extraordinary,'' J. Phys. A: Math. Theor. {\bf 57}, 285003 (2024).

\bibitem{OFC} Marian Anghel, private communication. 

\end{thebibliography}
\end{document}